\documentclass[aps,preprint,amsmath,amssymb,showpacs]{revtex4}
\usepackage{graphicx}

\begin{document}
\title{$WZ$ Production at $e\gamma$ Colliders and Anomalous Quartic  $WWZ\gamma$ Coupling}

\author{\.{I}. \c{S}ahin}
\email[]{isahin@science.ankara.edu.tr}
\affiliation{Department of Physics, Faculty of Sciences,
Ankara University, 06100 Tandogan, Ankara, Turkey}

\begin{abstract}
We investigate the constraints on the anomalous quartic
$W^{+}W^{-}Z\gamma$ gauge boson coupling through the process
$e^{-}\gamma\to \nu_{e}W^{-}Z$. Considering incoming beam
polarizations and the longitudinal and transverse polarization
states of the final W and Z boson we find 95\% confidence level
limits on the anomalous coupling parameter $a_{n}$ with an
integrated luminosity of 500 $fb^{-1}$ and $\sqrt{s}$=0.5, 1 TeV
energies. We show that initial beam and final state polarizations
improve the sensitivity to the anomalous coupling by up to factors
of 2 - 3.5 depending on the energy.
\end{abstract}

\pacs{12.15.Ji, 12.60.Cn, 13.88.+e}

\maketitle
\section{Introduction}
 The Standard Model (SM) has been a pillar of particle physics.
 It was subjected to many experimental tests but
SM has overcome so many of these experimental and theoretical
conflicts. In the recent experiments at CERN $e^{+}e^{-}$ collider
LEP and Fermilab Tevatron SM of electroweak interactions have been
tested with a good accuracy and the experimental results confirms
the $SU_{L}(2)\times U_{Y}(1)$ gauge structure of the SM.  However,
Higgs bosons have not been observed and one of the main goals of
future experiments is to pursue its trace. SM is largely silent on
the issue of the origin of the Higgs boson  and many physicists
believe that nature might use a more elegant way to accomplish
symmetry breaking and mass generation. These kind of considerations
motivate us to keep the trace of a more fundamental theory (new
physics) in which SM would be embedded.

Self-interactions of gauge bosons have not been tested with a good
accuracy and their precision measurements are in the scope of future
experiments. Precision measurements of these couplings
 will be the crucial test of the structure of the SM. Deviation
of the couplings from the expected values would indicate the
existence of new physics beyond the SM. In this work we analyzed
genuinely quartic $W^{+}W^{-}Z\gamma$ coupling which do not induce
new trilinear vertices. Genuine quartic couplings are contact
interactions, manifestations of the exchange of heavy particles.
They have different origins than anomalous trilinear couplings.
Trilinear couplings are form factors where heavy fields are
integrated out at the one-loop level. Therefore it is reasonable to
assume that quartic couplings are modified by genuine anomalous
interactions while the trilinear couplings are all given by their SM
values.

In writing effective operators associated to genuinely quartic
couplings we employ the formalism of \cite{eboli1}. Imposing
custodial $SU(2)_{Weak}$ symmetry and local $U(1)_{em}$ symmetry,
dimension 6 effective lagrangian for the $W^{+}W^{-}Z\gamma$
coupling is given by,

\begin{eqnarray}
{\cal L}_{n}&&=\frac{i\pi\alpha}{4\Lambda^{2}}a_{n}\epsilon_{ijk}
W_{\mu\alpha}^{(i)}W_{\nu}^{(j)}W^{(k)\alpha}F^{\mu\nu}
\end{eqnarray}
where $W^{(i)}$ is the $SU(2)_{Weak}$ triplet, and $F_{\mu\nu}$ and
$W_{\mu\alpha}^{(i)}$ are the electromagnetic and $SU(2)_{Weak}$
field strengths respectively. $a_{n}$ is the dimensionless anomalous
coupling constant. For sensitivity calculations to the anomalous
coupling we set the new physics energy scale $\Lambda$ to $M_{W}$.
The vertex function for
$W^{+}(p_{+}^{\mu})W^{-}(p_{-}^{\nu})Z(p_{1}^{\alpha})
\gamma(p_{2}^{\beta})$ generated from the effective lagrangian (1)
is given by

\begin{eqnarray}
i\frac{\pi\alpha}{4\cos\theta_{W}\Lambda^{2}}a_{n}\left[g_{\mu\alpha}\left[g_{\nu\beta}
p_{2}.(p_{1}-p_{+})-p_{2\nu}(p_{1}-p_{+})_{\beta}\right] \right.
\nonumber \\ \left.
-g_{\nu\alpha}\left[g_{\mu\beta}p_{2}.(p_{1}-p_{-})-p_{2\mu}(p_{1}-p_{-})_{\beta}\right]
\right. \nonumber \\ \left.
+g_{\mu\nu}\left[g_{\alpha\beta}p_{2}.(p_{+}-p_{-})-p_{2\alpha}(p_{+}-p_{-})_{\beta}\right]
 \right. \nonumber \\ \left. -p_{1\mu}(g_{\nu\beta} p_{2\alpha}-g_{\alpha\beta} p_{2\nu})
 +p_{1\nu}(g_{\mu\beta} p_{2\alpha}-g_{\alpha\beta} p_{2\mu})\right. \nonumber \\ \left.
-p_{-\alpha}(g_{\mu\beta} p_{2\nu}-g_{\nu\beta}p_{2\mu})
+p_{+\alpha}(g_{\nu\beta} p_{2\mu}-g_{\mu\beta}
p_{2\nu})\right.\nonumber \\ \left. -p_{+\nu}(g_{\alpha\beta}
p_{2\mu}-g_{\mu\beta}p_{2\alpha}) +p_{-\mu}(g_{\alpha\beta}
p_{2\nu}-g_{\nu\beta} p_{2\alpha})
 \right]
\end{eqnarray}

For a convention, we assume that all the momenta are incoming to the
vertex. It should be noted that lagrangian (1) represents only the
anomalous $W^{+}W^{-}Z\gamma$ coupling. In the cross section
calculations one should consider the lagrangian ${\cal L}={\cal
L}_{SM}+{\cal L}_{n}$ where ${\cal L}_{SM}$ is the SM lagrangian for
the vertex $W^{+}W^{-}Z\gamma$. Therefore within the SM, $a_{n}=0$.

CERN $e^{+}e^{-}$ collider LEP provide present collider limits on
anomalous quartic $W^{+}W^{-}Z\gamma$ coupling. At LEP the scaled
anomalous coupling $\frac{a_{n}}{\Lambda^{2}}$ is constrained by
analysing the process $e^{+}e^{-} \to W^{+}W^{-}\gamma$. This
process is sensitive to anomalous quartic gauge couplings in both
$W^{+}W^{-}Z\gamma$ and $W^{+}W^{-}\gamma\gamma$. Recent results
from L3, OPAL and DELPHI collaborations for $W^{+}W^{-}Z\gamma$
coupling are given by -0.14 $GeV^{-2} < \frac{a_{n}}{\Lambda^{2}} <$
0.13 $GeV^{-2}$, -0.16 $GeV^{-2} < \frac{a_{n}}{\Lambda^{2}} <$ 0.15
$GeV^{-2}$ and -0.18 $GeV^{-2} < \frac{a_{n}}{\Lambda^{2}} <$ 0.14
$GeV^{-2}$ at 95\% C.L. respectively \cite{l3}.

There have  been several studies in the literature for anomalous
quartic $W^{+}W^{-}Z\gamma$ coupling through the processes
 $e^{+}e^{-}\to W^{+}W^{-}Z, W^{+}W^{-}\gamma, W^{+}W^{-}(\gamma)\to 4f\gamma$
 \cite{denner},
 $e\gamma\to eW^{+}W^{-}, \nu_{e}W^{-}Z$ \cite{eboli1} and
$\gamma\gamma \to W^{+}W^{-}Z$ \cite{eboli2}. In the most of these
studies mentioned above future International Linear Collider (ILC)
and its $e\gamma$ and $\gamma\gamma$ modes have also been
considered. At the $e\gamma$ mode of ILC anomalous
$W^{+}W^{-}Z\gamma$ coupling appears in $eW^{+}W^{-}$ and
$\nu_{e}W^{-}Z$ production processes. As stated in ref.\cite{eboli1}
$\nu_{e}W^{-}Z$ production is much more sensitive to anomalous
coupling. Another advantage of $\nu_{e}W^{-}Z$ production is that it
isolates the $W^{+}W^{-}Z\gamma$ coupling. This feature is not seen
in any other processes mentioned above.

The LHC will start operating soon. A detailed analysis of bosonic
quartic couplings at the LHC via the processes $qq \to
qq\gamma\gamma$ and $qq \to qq\gamma Z(\to l^{+}l^{-})$ have been
done in ref.\cite{lietti}. The former process receives contributions
from the anomalous quartic couplings $ZZ\gamma\gamma$ and
$W^{+}W^{-}\gamma\gamma$ and the latter receives contributions from
$ZZZ\gamma$ and $W^{+}W^{-}Z\gamma$ \cite{lietti}. It was shown that
sensitivity bounds to the anomalous quartic $W^{+}W^{-}Z\gamma$
coupling through the process $qq \to qq\gamma Z(\to l^{+}l^{-})$ are
about the order of $O(10^{-6})$. However, the process $qq \to
qq\gamma Z(\to l^{+}l^{-})$ does not isolates $W^{+}W^{-}Z\gamma$
coupling and the bounds were obtained under the assumption that only
one anomalous coupling is different from zero \cite{lietti}.

In this work  we consider the process $e\gamma\to \nu_{e}W^{-}Z$
to investigate $W^{+}W^{-}Z\gamma$ coupling. This process was
analyzed in ref.\cite{eboli1} with unpolarized beams. We take
account of incoming beam polarizations and also the longitudinal
and transverse polarization states of the final gauge bosons in
the cross section calculations to improve the bounds, assuming the
polarization of W and Z can be measured \cite{opal2}.

\section{Cross sections for polarized beams}

The process $e\gamma\to \nu_{e}W^{-}Z$ takes part as a subprocess
in $e^{+}e^{-}$ collision. Real gamma beam which enters the
subprocess is obtained by Compton backscattering of laser light
off linear electron or positron beam where most of the photons are
produced at the high energy region.

The spectrum of backscattered photons in connection with
helicities of initial laser photon and electron is \cite{ginzb1}:

\begin{eqnarray}
f_{\gamma/e}(y)={{1}\over{g(\zeta)}}[1-y+{{1}\over{1-y}}
-{{4y}\over{\zeta(1-y)}}+{{4y^{2}}\over {\zeta^{2}(1-y)^{2}}}+
\lambda_{0}\lambda_{e} r\zeta (1-2r)(2-y)]
\end{eqnarray}

where

\begin{eqnarray}
g(\zeta)=&&g_{1}(\zeta)+
\lambda_{0}\lambda_{e}g_{2}(\zeta) \nonumber\\
g_{1}(\zeta)=&&(1-{{4}\over{\zeta}}
-{{8}\over{\zeta^{2}}})\ln{(\zeta+1)}
+{{1}\over{2}}+{{8}\over{\zeta}}-{{1}\over{2(\zeta+1)^{2}}} \\
g_{2}(\zeta)=&&(1+{{2}\over{\zeta}})\ln{(\zeta+1)}
-{{5}\over{2}}+{{1}\over{\zeta+1}}-{{1}\over{2(\zeta+1)^{2}}}
\end{eqnarray}

Here $r=y/[\zeta(1-y)]$ and $\zeta=4E_{e}E_{0}/M_{e}^{2}$. $E_{0}$
and $\lambda_{0}$ are the energy and helicity of initial laser
photon and $E_{e}$ and $\lambda_{e}$ are the energy and the
helicity of initial electron beam before Compton backscattering.
$y$ is the fraction which represents the ratio between the
scattered photon and initial electron energy for the backscattered
photons moving along the initial electron direction. Maximum value
of $y$ reaches 0.83 when $\zeta=4.8$ in which the backscattered
photon energy is maximized without spoiling the luminosity.
Backscattered photons are not in fixed helicity states their
helicities are described by a distribution :

\begin{eqnarray}
\xi(E_{\gamma},\lambda_{0})={{\lambda_{0}(1-2r)
(1-y+1/(1-y))+\lambda_{e} r\zeta[1+(1-y)(1-2r)^{2}]}
\over{1-y+1/(1-y)-4r(1-r)-\lambda_{e}\lambda_{0}r\zeta
(2r-1)(2-y)}}
\end{eqnarray}

The helicity dependent differential cross section for the
subprocess can be connected to initial laser photon helicity
$\lambda_{0}$ and initial electron beam polarization $P_{e}$
through the formula,

\begin{eqnarray}
&&d\hat{\sigma}(\lambda_{0},P_{e};\lambda_{W},\lambda_{Z})\nonumber
\\ &&=\frac{1}{4}(1-P_{e})\left[(1+\xi(E_{\gamma},\lambda_{0}))
d\hat{\sigma}(+,L;\lambda_{W},\lambda_{Z})+(1-\xi(E_{\gamma},\lambda_{0}))
d\hat{\sigma}(-,L;\lambda_{W},\lambda_{Z})\right]\nonumber \\
&&+\frac{1}{4}(1+P_{e})\left[(1+\xi(E_{\gamma},\lambda_{0}))
d\hat{\sigma}(+,R;\lambda_{W},\lambda_{Z})+(1-\xi(E_{\gamma},\lambda_{0}))
d\hat{\sigma}(-,R;\lambda_{W},\lambda_{Z})\right]
\end{eqnarray}

Here
$d\hat{\sigma}(\lambda_{\gamma},\sigma;\lambda_{W},\lambda_{Z})$
is the helicity dependent differential cross section in the
helicity eigenstates; $\sigma: L,R$, $\lambda_{\gamma}=+,-$ and
$\lambda_{W},\lambda_{Z}=+,-,0$. It should be noted that $P_{e}$
and $\lambda_{e}$ refer to different beams. $P_{e}$ is the
electron beam polarization which enters the subprocess but
$\lambda_{e}$ is the polarization of initial electron beam before
Compton backscattering. The integrated cross section can be
obtained by integrating the cross section (7) for the subprocess
over the backscattered photon spectrum.

The process $e^{-}\gamma\to \nu_{e}W^{-}Z$ is described by nine
tree-level diagrams. Only the t-channel W exchange diagram
contains anomalous $W^{+}W^{-}Z\gamma$ coupling. The helicity
amplitudes have been calculated using vertex amplitude techniques
derived in ref.\cite{maina} and the phase space integrations have
been performed by GRACE \cite{grace} which uses a Monte Carlo
routine.

In our calculations we accept that initial electron beam
polarizability is $|\lambda_{e}|, |P_{e}|$=0.8. To see the
influence of initial beam polarization, energy distributions of
backscattered photons $f_{\gamma /e}$ are plotted for
$\lambda_{e}\lambda_{0}$=0, -0.8 and +0.8 in Fig. 1. We see from
the figure that backscattered photon distribution is very low at
high energies in $\lambda_{e}\lambda_{0}$=+0.8. Therefore we will
only consider the case $\lambda_{e}\lambda_{0}< 0$ in the cross
section calculations. Moreover the Feynman diagram containing
anomalous $W^{+}W^{-}Z\gamma$ coupling is a W exchange diagram
with a $We\nu_{e}$ vertex. Due to V-A structure of the $We\nu_{e}$
vertex,
$d\hat{\sigma}(\lambda_{\gamma},L;\lambda_{W},\lambda_{Z})$ is
more sensitive to anomalous coupling than
$d\hat{\sigma}(\lambda_{\gamma},R;\lambda_{W},\lambda_{Z})$. So we
will consider the case in which $P_{e}$=-0.8 (see eq. (7)).

One can see from Fig.~\ref{fig2}-~\ref{fig3} the influence of the
final state polarizations on the deviations of the total cross
sections from their SM value for initial beam polarizations
$(\lambda_{e},\lambda_{0},P_{e})=(-0.8,1,-0.8)$ and
$(\lambda_{e},\lambda_{0},P_{e})=(0.8,-1,-0.8)$. In these figures
TR and LO stand for "transverse" and "longitudinal" respectively.
Transverse polarization configuration of the final bosons are
almost insensitive to anomalous coupling. Therefore we omit them
in the figures. It is clear from Fig.~\ref{fig2}-~\ref{fig3} that
longitudinally polarized cross sections are sensitive to anomalous
coupling. For instance in Fig.~\ref{fig2} cross section at the
polarization configuration $(\lambda_{W},\lambda_{Z})$=(LO,LO)
increases by a factor of 3.6 as $a_{n}$ increases from 0 to 1. But
this increment is only a factor of 1.2 in the unpolarized case.

In Fig.~\ref{fig4} longitudinally polarized total cross sections
are plotted as a function of anomalous coupling $a_{n}$ for
different initial beam polarizations. Center of mass energy of the
$e^{+}e^{-}$ system is $\sqrt{s}=0.5$ TeV. We see from the
Fig.~\ref{fig4} the effect of initial beam polarizations on the
deviations of cross sections from the SM.

\section{Angular correlations for final state fermions}
Angular distributions of $W^{-}$ and $Z$ decay products have clear
correlations with the helicity states of these final state gauge
bosons. Therefore in principle, polarization states of final $W^{-}$
and $Z$ boson can be determined by measuring the angular
distributions of $W^{-}$ and $Z$ decay products. This kind of
treatment was done in reference \cite{hagiwara} for final state
$W^{-}$ and $W^{+}$ bosons. Let us consider the differential cross
section for the complete process,

\begin{eqnarray}
 e^{-}(k_{1},\sigma)+\gamma(k_{2},\lambda_{\gamma}) \to \nu_{e}(q_{1},\bar{\sigma})+
 W^{-}(q_{2},\lambda_{W})+Z(q_{3},\lambda_{Z}) \nonumber \\
W^{-}(q_{2},\lambda_{W}) \to f_{1}(p_{1},\sigma_{1})
\bar{f_{2}}(p_{2},\sigma_{2}) \nonumber \\Z(q_{3},\lambda_{Z}) \to
f_{3}(p_{3},\sigma_{3}) \bar{f_{4}}(p_{4},\sigma_{4})
\end{eqnarray}

with massless fermions
$f_{1}$,$\bar{f_{2}}$,$f_{3}$,$\bar{f_{4}}$. Here $\sigma$ and
$\lambda_{\gamma}$ are the incoming electron and photon
helicities; $\bar{\sigma}$, $\lambda_{W}$ and $\lambda_{Z}$ are
the outgoing $\nu_{e}$, $W^{-}$ and $Z$ helicities. $\sigma_{i}$
represent the helicities of final fermions $f_{i}$ or
$\bar{f_{i}}$.

The full amplitude can be expressed as follows:

\begin{eqnarray}
M(k_{1},\sigma;k_{2},\lambda_{\gamma};q_{1},\bar{\sigma};p_{i},\sigma_{i})
=D_{W}(q_{2}^{2})D_{Z}(q_{3}^{2})\sum_{\lambda_{W}}\sum_{\lambda_{Z}}
M_{1}(k_{1},\sigma;k_{2},\lambda_{\gamma};q_{1},\bar{\sigma};q_{2},\lambda_{W};q_{3},\lambda_{Z})
\nonumber
\\ \times M_{2}(q_{2},\lambda_{W};p_{1},\sigma_{1};p_{2},\sigma_{2})
\times M_{3}(q_{3},\lambda_{Z};p_{3},\sigma_{3};p_{4},\sigma_{4})
\end{eqnarray}

where
$M_{1}(k_{1},\sigma;k_{2},\lambda_{\gamma};q_{1},\bar{\sigma};q_{2},\lambda_{W};q_{3},\lambda_{Z})$
is the production amplitude; helicity amplitudes for $e^{-}\gamma
\to \nu_{e}W^{-}Z$ with on-shell $W^{-}$ and $Z$ boson.
$M_{2}(q_{2},\lambda_{W};p_{1},\sigma_{1};p_{2},\sigma_{2})$ and
$M_{3}(q_{3},\lambda_{Z};p_{3},\sigma_{3};p_{4},\sigma_{4})$ are
the decay amplitudes of $W^{-}$ and $Z$ boson to fermions.
$D_{W}(q_{2}^{2})$ and $D_{Z}(q_{3}^{2})$ are the Breit-Wigner
propagator factors for $W^{-}$ and $Z$ bosons.

In this paper we consider lepton decay channel of final state
bosons. Therefore $f_{1}$,$\bar{f_{2}}$,$f_{3}$,$\bar{f_{4}}$ are
leptons. $M_{2}$ and $M_{3}$ decay amplitudes are most simply
expressed in the rest frames of $W^{-}$ and $Z$ respectively. In
the $W^{-}$ rest frame, four-momenta of $W^{-}$ decay products
$f_{1}$ and $\bar{f_{2}}$ can be parametrized as

\begin{eqnarray}
p_{1}^{\mu}=&&\frac{m_{W}}{2}(1,sin\theta cos\phi,sin\theta
sin\phi,cos\theta) \nonumber \\
p_{2}^{\mu}=&&\frac{m_{W}}{2}(1,-sin\theta cos\phi,-sin\theta
sin\phi,-cos\theta)
\end{eqnarray}

where $\theta$ and $\phi$ are polar and azimuthal angles in the
$W^{-}$ rest frame with respect to z-axis defined to be the
$W^{-}$ boson direction in the $e^{+}e^{-}$ center of mass frame
(lab. frame). $W^{-}$ rest frame is defined by a boost of the
$e^{+}e^{-}$ center of mass frame along the z-axis. In this rest
frame $M_{2}$ decay amplitude is given by

\begin{eqnarray}
M_{2}=&&\frac{g_{W}}{\sqrt{2}}m_{W}\delta_{\sigma_{1},-}\delta_{\sigma_{2},+}l_{\lambda_{W}}
\end{eqnarray}

with

\begin{eqnarray}
(l_{-},l_{0},l_{+})=&&(\frac{1}{\sqrt{2}}(1+cos\theta)e^{-i\phi},-sin\theta,\frac{1}{\sqrt{2}}(1-cos\theta)e^{i\phi})
\end{eqnarray}

In the $Z$ rest frame we parametrize four-momenta of $f_{3}$ and
$\bar{f_{4}}$ as

\begin{eqnarray}
p_{3}^{\mu}=&&\frac{m_{Z}}{2}(1,sin\bar{\theta}
cos\bar{\phi},sin\bar{\theta}
sin\bar{\phi},cos\bar{\theta}) \nonumber \\
p_{4}^{\mu}=&&\frac{m_{Z}}{2}(1,-sin\bar{\theta}
cos\bar{\phi},-sin\bar{\theta} sin\bar{\phi},-cos\bar{\theta})
\end{eqnarray}

where $\bar{\theta}$ and $\bar{\phi}$ are polar and azimuthal
angles in the $Z$ rest frame with respect to $\bar{z}$-axis
defined to be the $Z$ boson direction in the $e^{+}e^{-}$ center
of mass frame. In the $Z$ rest frame $M_{3}$ decay amplitude is
given by

\begin{eqnarray}
M_{3}=&&m_{Z}\left[-g_{L}\delta_{\sigma_{3},-}\delta_{\sigma_{4},+}\bar{l}_{L}(\lambda_{Z})
+g_{R}\delta_{\sigma_{3},+}\delta_{\sigma_{4},-}\bar{l}_{R}(\lambda_{Z})\right]
\end{eqnarray}

with

\begin{eqnarray}
(\bar{l}_{L}(-),\bar{l}_{L}(0),\bar{l}_{L}(+))=&&(\frac{1}{\sqrt{2}}
(1+cos\bar{\theta})e^{-i\bar{\phi}},-sin\bar{\theta},\frac{1}{\sqrt{2}}(1-cos\bar{\theta})e^{i\bar{\phi}})
\nonumber\\
(\bar{l}_{R}(-),\bar{l}_{R}(0),\bar{l}_{R}(+))=&&(\frac{1}{\sqrt{2}}
(1-cos\bar{\theta})e^{-i\bar{\phi}},sin\bar{\theta},\frac{1}{\sqrt{2}}(1+cos\bar{\theta})e^{i\bar{\phi}})\\
g_{L}=&&g_{Z}\frac{(C_{V}+C_{A})}{2}
,\,\,\,\,\,\,\,\,g_{R}=g_{Z}\frac{(C_{V}-C_{A})}{2}
\end{eqnarray}

where $C_{V}$ and $C_{A}$ are usual vector and axial vector
couplings.

Polarization summed squared matrix elements are given by

\begin{eqnarray}
\sum_{\sigma,\lambda_{\gamma},\bar{\sigma},\sigma_{i}}|M(k_{1},\sigma;k_{2},\lambda_{\gamma};
q_{1},\bar{\sigma};p_{i},\sigma_{i})|^{2}=|D_{W}(q_{2}^{2})|^{2}|D_{Z}(q_{3}^{2})|^{2}P_{\lambda_{W}^{\prime}
\lambda_{Z}^{\prime}}^{\lambda_{W}\lambda_{Z}}D_{\lambda_{W}^{\prime}}^{\lambda_{W}}\bar{D}_{
\lambda_{Z}^{\prime}}^{\lambda_{Z}}
\end{eqnarray}

In this equation summation over repeated indices
$(\lambda_{W},\lambda_{W}^{\prime},\lambda_{Z},\lambda_{Z}^{\prime})=+,-,0$
is implied. $P_{\lambda_{W}^{\prime}
\lambda_{Z}^{\prime}}^{\lambda_{W}\lambda_{Z}}$ is the production
tensor and $D_{\lambda_{W}^{\prime}}^{\lambda_{W}},\bar{D}_{
\lambda_{Z}^{\prime}}^{\lambda_{Z}}$ are the decay tensors for W
and Z boson respectively. They are defined by

\begin{eqnarray}
P_{\lambda_{W}^{\prime}
\lambda_{Z}^{\prime}}^{\lambda_{W}\lambda_{Z}}=&&\sum_{\sigma,\lambda_{\gamma},\bar{\sigma}}
M_{1}(k_{1},\sigma;k_{2},\lambda_{\gamma};q_{1},\bar{\sigma};q_{2},\lambda_{W};q_{3},\lambda_{Z})\nonumber
\\
&&\times
M_{1}^{\star}(k_{1},\sigma;k_{2},\lambda_{\gamma};q_{1},\bar{\sigma};
q_{2},\lambda_{W}^{\prime};q_{3},\lambda_{Z}^{\prime})
\\D_{\lambda_{W}^{\prime}}^{\lambda_{W}}=&&\sum_{\sigma_{1},\sigma_{2}}
M_{2}(q_{2},\lambda_{W};p_{1},\sigma_{1};p_{2},\sigma_{2})
M_{2}^{\star}(q_{2},\lambda_{W}^{\prime};p_{1},\sigma_{1};p_{2},\sigma_{2})
\\\bar{D}_{\lambda_{Z}^{\prime}}^{\lambda_{Z}}=&&\sum_{\sigma_{3},\sigma_{4}}
M_{3}(q_{3},\lambda_{Z};p_{3},\sigma_{3};p_{4},\sigma_{4})
M_{3}^{\star}(q_{3},\lambda_{Z}^{\prime};p_{3},\sigma_{3};p_{4},\sigma_{4})
\end{eqnarray}

Now let us write down the differential cross section :

\begin{eqnarray}
d\sigma=&&\frac{1}{2s}|M|^{2}\frac{d^{3}q_{1}}{(2\pi)^{3}2E_{q_{1}}}
\frac{d^{3}p_{1}}{(2\pi)^{3}2E_{1}}
\frac{d^{3}p_{2}}{(2\pi)^{3}2E_{2}}
\frac{d^{3}p_{3}}{(2\pi)^{3}2E_{3}}
\frac{d^{3}p_{4}}{(2\pi)^{3}2E_{4}}\nonumber \\
 &&\times(2\pi)^{4}\delta^{4}(k_{1}+k_{2}-q_{1}-p_{1}-p_{2}-p_{3}-p_{4})
\end{eqnarray}

Using narrow width approximation it is straightforward to express
the differential cross section as

\begin{eqnarray}
d\sigma=&&\frac{1}{2s}(2\pi)^{4}\delta^{4}(k_{1}+k_{2}-q_{1}-q_{2}-q_{3})\frac{\pi^{2}}
{2^{6}(2\pi)^{6}\Gamma_{W}\Gamma_{Z}m_{W}m_{Z}}\nonumber \\
&&\times
P_{\lambda_{W}^{\prime}\lambda_{Z}^{\prime}}^{\lambda_{W}\lambda_{Z}}
D_{\lambda_{W}^{\prime}}^{\lambda_{W}}\bar{D}_{\lambda_{Z}^{\prime}}^{\lambda_{Z}}
\frac{d^{3}q_{1}}{(2\pi)^{3}2E_{q_{1}}}\frac{d^{3}q_{2}}{(2\pi)^{3}2E_{q_{2}}}
\frac{d^{3}q_{3}}{(2\pi)^{3}2E_{q_{3}}} \,\,dcos\theta d\phi \,\,
dcos\bar{\theta} d\bar{\phi}
\end{eqnarray}

After integration over azimuthal angles $\phi$ and $\bar{\phi}$
interference terms will vanish and only the diagonal terms
$\lambda_{W}=\lambda_{W}^{\prime}$ and
$\lambda_{Z}=\lambda_{Z}^{\prime}$ will survive. It is now
 straightforward to write differential cross section in the form:

\begin{eqnarray}
d\sigma=d\sigma_{1}(\lambda_{W},\lambda_{Z})d_{\lambda_{W}}^{\lambda_{W}}\bar{d}_{\lambda_{Z}}^{\lambda_{Z}}
\frac{9}{32(C_{V}^{2}+C_{A}^{2})}B(W \to l\bar{\nu}_{l})B(Z \to
l^{+}l^{-})dcos\theta dcos\bar{\theta}
\end{eqnarray}

Here $d\sigma_{1}(\lambda_{W},\lambda_{Z})$ is the helicity
dependent production cross section, $B(W \to l\bar{\nu}_{l})$ and
$B(Z \to l^{+}l^{-})$ are the branching ratios of W and Z boson to
leptons. The matrices $d_{\lambda_{W}}^{\lambda_{W}}$ and
$\bar{d}_{\lambda_{Z}}^{\lambda_{Z}}$ are related to the diagonal
elements of decay tensors (19-20) as

\begin{eqnarray}
d_{\lambda_{W}}^{\lambda_{W}}=&&l_{\lambda_{W}}l^{\star}_{\lambda_{W}}\\
\bar{d}_{\lambda_{Z}}^{\lambda_{Z}}=&&\left[(C_{V}+C_{A})^{2}\bar{l}_{L}(\lambda_{Z})\bar{l}^{\star}_{L}(\lambda_{Z})
+(C_{V}-C_{A})^{2}\bar{l}_{R}(\lambda_{Z})\bar{l}^{\star}_{R}(\lambda_{Z})\right]
\end{eqnarray}

It is difficult to identify nine different polarization
configurations of the production cross section but it is sensible
to claim that longitudinal (LO) and transverse (TR) polarizations
can be identified \cite{hagiwara}. Thus we define the following
cross sections:

\begin{eqnarray}
d\sigma_{1}(TR,TR)=&&\sum_{\lambda_{W}=+,-}\sum_{\lambda_{Z}=+,-}d\sigma_{1}(\lambda_{W},\lambda_{Z})\\
d\sigma_{1}(LO,LO)=&&d\sigma_{1}(0,0)\\
d\sigma_{1}(TR,LO)=&&\sum_{\lambda_{W}=+,-}d\sigma_{1}(\lambda_{W},0)\\
d\sigma_{1}(LO,TR)=&&\sum_{\lambda_{Z}=+,-}d\sigma_{1}(0,\lambda_{Z})\\
\end{eqnarray}

and

\begin{eqnarray}
d\sigma_{1}(TR,unpol)=d\sigma_{1}(TR,TR)+d\sigma_{1}(TR,LO)\\
d\sigma_{1}(LO,unpol)=d\sigma_{1}(LO,TR)+d\sigma_{1}(LO,LO)\\
d\sigma_{1}(unpol,TR)=d\sigma_{1}(TR,TR)+d\sigma_{1}(LO,TR)\\
d\sigma_{1}(unpol,LO)=d\sigma_{1}(TR,LO)+d\sigma_{1}(LO,LO)
\end{eqnarray}

For fixed W and Z helicities above cross sections can be obtained
from a fit to polar angle distributions of the W and Z decay
products in the W and Z rest frames . To be precise for
$\lambda_{W}=\pm1,0$ polarization states of final W, production
cross sections $d\sigma_{1}(\pm,\lambda_{Z})$ and
$d\sigma_{1}(0,\lambda_{Z})$ can be obtained from a fit to
$d_{+}^{+}, d_{-}^{-}$ and $d_{0}^{0}$ distributions in the W rest
frame (eqn.(23)). Similarly production cross sections
$d\sigma_{1}(\lambda_{W},\pm)$ and $d\sigma_{1}(\lambda_{W},0)$ can
be obtained from a fit to $\bar{d}_{+}^{+}, \bar{d}_{-}^{-}$ and
$\bar{d}_{0}^{0}$ distributions in the Z rest frame. In
Fig.~\ref{fig5} and Fig.~\ref{fig6} $d_{\lambda_{W}}^{\lambda_{W}}$
and $\bar{d}_{\lambda_{Z}}^{\lambda_{Z}}$ distributions are plotted
for various polarization states of final W and Z boson. As can be
seen from these figures longitudinal (LO) and transverse (TR)
distributions are well separated from each other.

There have been several experimental studies in the literature for
the measurement of W polarization \cite{opal2}. It is reasonable to
assume that Z polarization can be accessible in a similar manner. At
lepton colliders systematic uncertainties are expected to be lower
than hadronic colliders. For this reason in our calculations we will
ignore the uncertainties associated to the determination of the
polarizations of final state gauge bosons.

\section{limits on the Anomalous Coupling Parameter}

A detailed investigation of the anomalous couplings requires a
statistical analysis. To this purpose we have obtained 95\% C.L.
limits on the anomalous coupling parameter $a_{n}$ using
$\chi^{2}$ analysis at $\sqrt{s}=0.5, 1$ TeV and integrated
luminosity $L_{int}=500$ $fb^{-1}$ without systematic errors. The
number of events are given as $N=AL_{int}\sigma B_{W}B_{Z}$ where
$A$ is the overall acceptance and $B_{W}$ and $B_{Z}$ are the
branching ratios of W and Z boson for leptonic channel.

The limits for the anomalous $W^{+}W^{-}Z\gamma$ coupling are
given on Table \ref{tab1} for unpolarized initial beams and
unpolarized, transverse and longitudinal polarization states of
final W and Z boson with the acceptance $A=0.85$. One can see from
Table \ref{tab1} that polarization configuration
$(\lambda_{W},\lambda_{Z})$=(LO, TR+LO) is most sensitive to
anomalous coupling at $\sqrt{s}=0.5$ TeV. This configuration
improves the limits by a factor of 1.3. But at $\sqrt{s}=1$ TeV
polarization configuration $(\lambda_{W},\lambda_{Z})$=(LO, LO) is
the most sensitive and improves the limits by a factor of 2.

On Table \ref{tab2} and  \ref{tab3}  the initial beam polarizations
are also taken into account. One can see from Table \ref{tab2} that
polarization configurations
$(\lambda_{0},\lambda_{e},\lambda_{W},\lambda_{Z})$=(1, -0.8, LO,
LO) or (1, -0.8, LO, TR+LO) improves the limits by a factor of 2.
Increase in energy highly improves the limits. At $\sqrt{s}=1$ TeV
the most sensitive polarization configuration is
$(\lambda_{0},\lambda_{e},\lambda_{W},\lambda_{Z})$=(1, -0.8, LO,
LO) and this configuration improves the limits by a factor of 3.5.

Anomalous $W^{+}W^{-}Z\gamma$ coupling was studied in
ref.\cite{eboli1} through the same process $e^{-}\gamma\to
\nu_{e}W^{-}Z$ with unpolarized beams. Using statistical
significance authors set $3\sigma$ bound of (-1.2, 0.74) on the
anomalous $W^{+}W^{-}Z\gamma$ coupling parameter $a_{n}$ with an
integrated luminosity of 10 $fb^{-1}$ and $\sqrt{s}$=0.5 TeV energy.
In order to compare our results with the results of
ref.\cite{eboli1} we have calculated $3\sigma$ significance bounds
with an integrated luminosity of 10 $fb^{-1}$ and $\sqrt{s}$=0.5 TeV
energy. For unpolarized beams we have confirmed the result of
ref.\cite{eboli1}. The most sensitive results are obtained at the
polarization configurations
$(\lambda_{0},\lambda_{e},\lambda_{W},\lambda_{Z})$=(1, -0.8, LO,
LO) and (1, -0.8, LO, TR+LO). $3\sigma$ significance bounds for the
polarization configurations
$(\lambda_{0},\lambda_{e},\lambda_{W},\lambda_{Z})$= (1, -0.8, LO,
LO) and (1, -0.8, LO, TR+LO) are given by (-0.59, 0.42) and (-0.63,
0.38) respectively. Therefore polarization improves the significance
bounds of $a_{n}$ approximately a factor of 1.92 for integrated
luminosity $L_{int}= 10 fb^{-1}$ and $\sqrt{s}$=0.5 TeV.

The $e\gamma$ mode of ILC with luminosity $L_{int}=500$ $fb^{-1}$
probes the anomalous $W^{+}W^{-}Z\gamma$ coupling with far better
sensitivity than the present collider LEP2 experiments. It improves
the sensitivity limits by up to a factor of $10^{4}$ with respect to
LEP2. This is comparable with the limits which are expected to be
obtained at CERN LHC \cite{lietti}. One prominent advantage of the
process $e^{-}\gamma\to \nu_{e}W^{-}Z$ is that it isolates anomalous
$W^{+}W^{-}Z\gamma$ coupling. It provides us the opportunity to
study $W^{+}W^{-}Z\gamma$ coupling independent from $ZZZ\gamma$ as
well as $ZZ\gamma\gamma$ and $W^{+}W^{-}\gamma\gamma$. In
conclusion, experiments with polarized $e^{+}e^{-}$ beams and final
state polarizations leads to a significant improvement in the
sensitivity limits. Although the SM cross sections in the
longitudinal polarization configurations of the final W and Z boson
are small, sensitivity limits are better than the transverse
polarization case.

\pagebreak

\pagebreak

\begin{figure}
\includegraphics{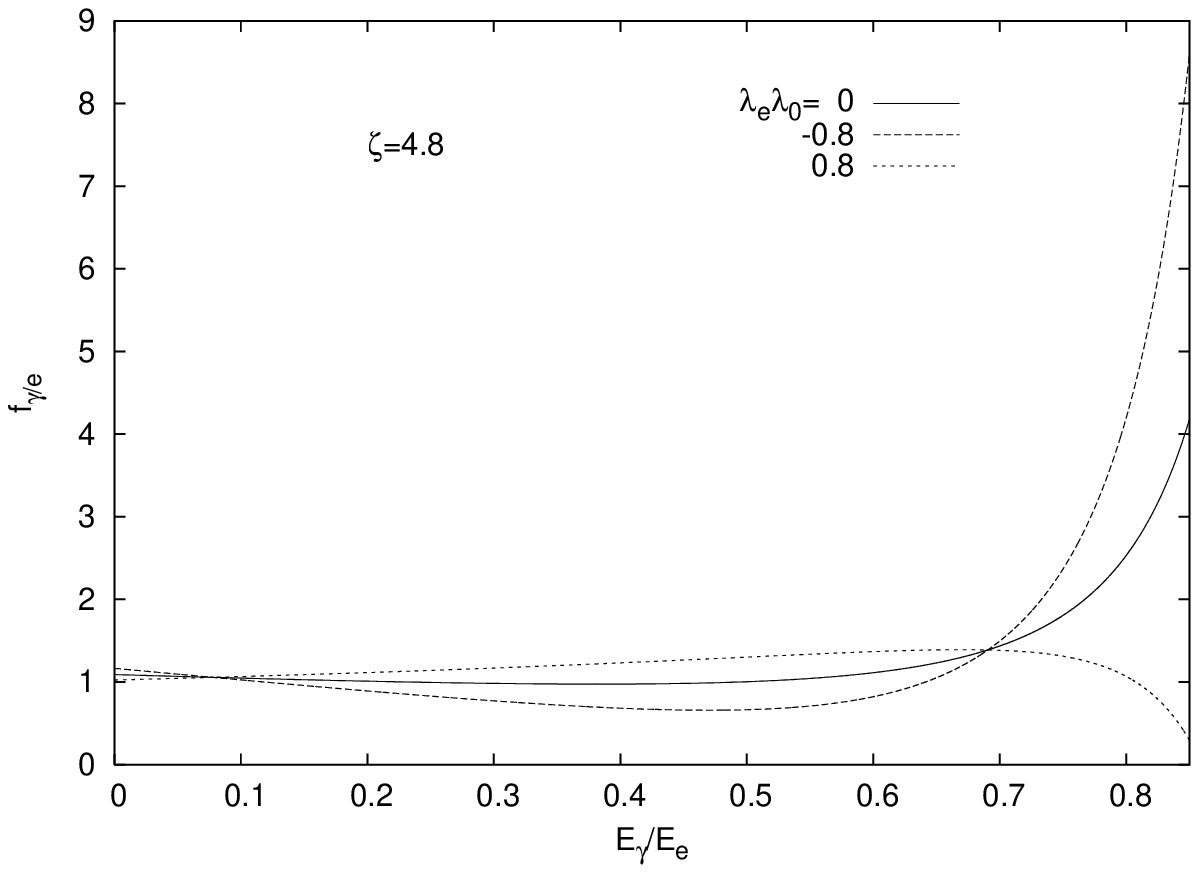}
\caption{Energy distribution of backscattered photons for
$\lambda_{e}\lambda_{0}=0, -0.8, 0.8.$ \label{fig1}}
\end{figure}

\begin{figure}
\includegraphics{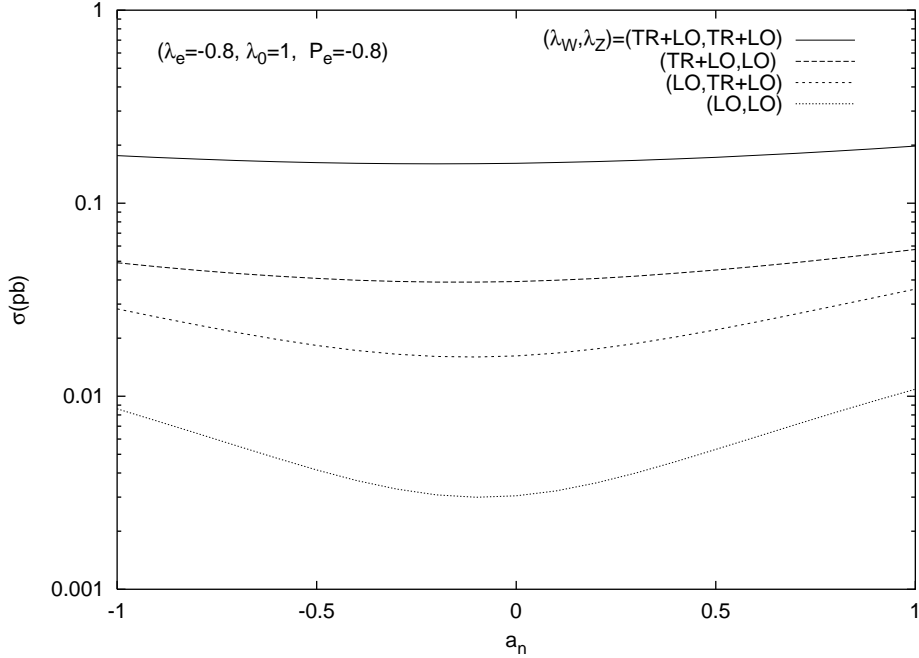}
\caption{The integrated total cross section of $e^{-}\gamma\to
\nu_{e}W^{-}Z$ as a function of anomalous coupling $a_{n}$ for
initial beam polarization
$(\lambda_{e},\lambda_{0},P_{e})=(-0.8,1,-0.8)$ and final state
polarizations stated on the figure. $\sqrt{s}=0.5$ TeV.
\label{fig2}}
\end{figure}

\begin{figure}
\includegraphics{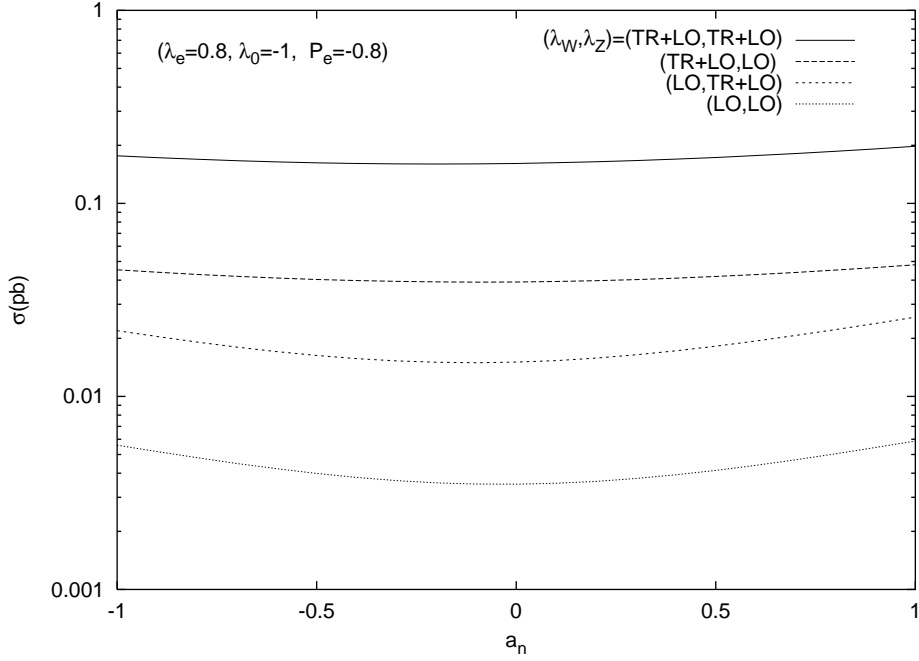}
\caption{The same as Fig. 2 but for
$(\lambda_{e},\lambda_{0},P_{e})=(0.8,-1,-0.8)$ \label{fig3}}
\end{figure}

\begin{figure}
\includegraphics{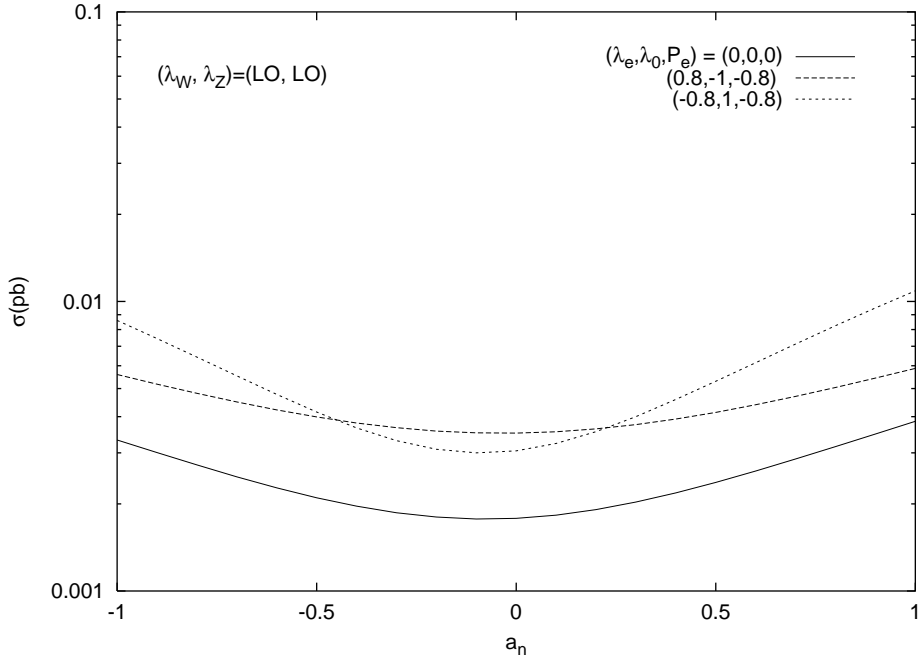}
\caption{The integrated total cross section of $e^{-}\gamma\to
\nu_{e}W^{-}Z$ as a function of anomalous coupling $a_{n}$ for
final state polarization configuration
$(\lambda_{W},\lambda_{Z})=(LO,LO)$. The legends are for initial
beam polarizations. $\sqrt{s}=0.5$ TeV. \label{fig4}}
\end{figure}

\begin{figure}
\includegraphics{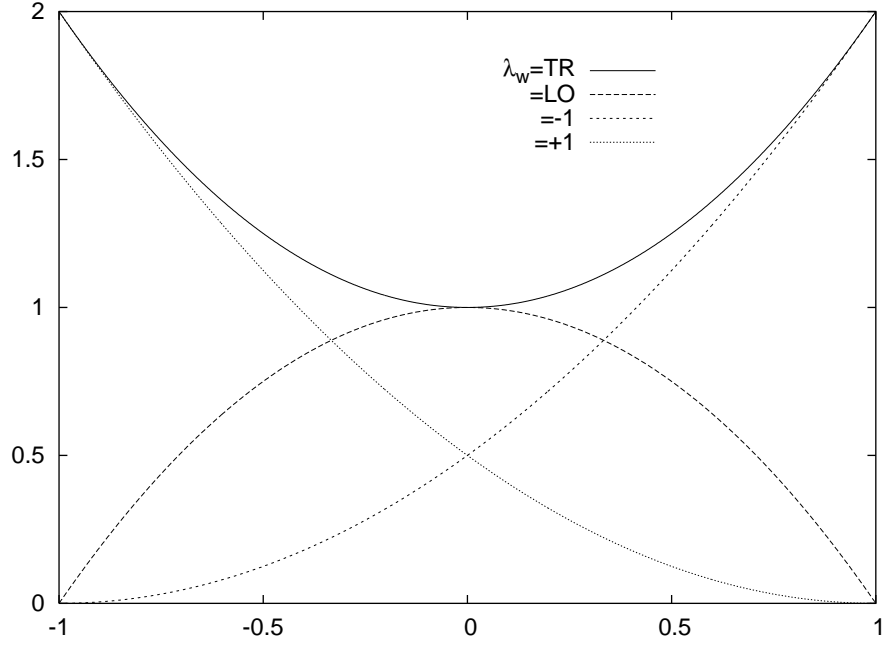}
\caption{$d_{\lambda_{W}}^{\lambda_{W}}$ versus $cos\theta$. The
legends are for various polarization states of the final W boson.
\label{fig5}}
\end{figure}

\begin{figure}
\includegraphics{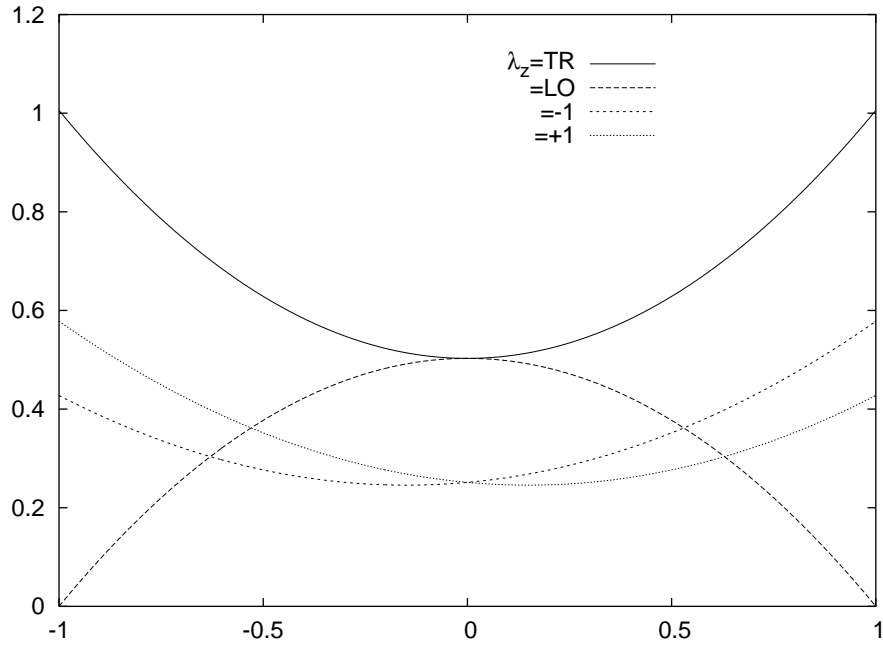}
\caption{$\bar{d}_{\lambda_{Z}}^{\lambda_{Z}}$ versus
$cos\bar{\theta}$. The legends are for various polarization states
of the final Z boson. \label{fig6}}
\end{figure}

\begin{table}
\caption{Sensitivity of the  $e\gamma $ collision to $WWZ\gamma$
couplings at 95\% C.L. for $\sqrt{s}=0.5, 1$ TeV and $L_{int}=500$
$fb^{-1}$. The initial beams are unpolarized. The effects of final
state W and Z boson polarizations are shown in each
row.\label{tab1}}
\begin{ruledtabular}
\begin{tabular}{cccc}
$\sqrt{s}$  TeV& $\lambda_{W}$& $\lambda_{Z}$& $a_{n}$ \\
\hline
  0.5 & TR+LO &TR+LO &-0.95, 0.50\\
  0.5 & LO &TR+LO &-0.70, 0.40\\
  0.5 & TR &TR+LO &-1.42, 0.75\\
  0.5 & TR+LO &LO &-0.87, 0.55\\
  0.5 & TR+LO &TR &-1.25, 0.70\\
  0.5 & LO &LO &-0.70, 0.55\\
  0.5 & TR &TR &-2.60, 1.20\\
  0.5 & LO &TR &-0.80, 0.49\\
  0.5 & TR &LO &-1.10, 0.70\\
  \hline
  1 & TR+LO &TR+LO &-0.12, 0.09\\
  1 & LO &TR+LO &-0.07, 0.06\\
  1 & TR &TR+LO &-0.22, 0.17\\
  1 & TR+LO &LO &-0.10, 0.09\\
  1 & TR+LO &TR &-0.19, 0.14\\
  1 & LO &LO &-0.06, 0.05\\
  1 & TR &TR &-0.70, 0.40\\
  1 & LO &TR &-0.10, 0.08\\
  1 & TR &LO &-0.16, 0.13\\
\end{tabular}
\end{ruledtabular}
\end{table}

\begin{table}
\caption{Sensitivity of the  $e\gamma $ collision to $WWZ\gamma$
couplings at 95\% C.L. for $\sqrt{s}=0.5$ TeV and $L_{int}=500$
$fb^{-1}$. The effects of final state W, Z boson and initial beam
polarizations are shown in each row.\label{tab2}}
\begin{ruledtabular}
\begin{tabular}{cccccc}
$\lambda_{0}$&$\lambda_{e}$ & $P_{e} $& $\lambda_{W}$&
 $\lambda_{Z}$& $a_{n}$ \\
\hline
 1 & -0.8 & -0.8 &  TR+LO& TR+LO &-0.75, 0.34  \\
1 & -0.8 & -0.8 &  LO& TR+LO &-0.50, 0.26  \\
1 & -0.8 & -0.8 &  TR& TR+LO &-1.20, 0.52  \\
1 & -0.8 & -0.8 &  TR+LO& LO &-0.65, 0.35  \\
1 & -0.8 & -0.8 &  TR+LO& TR &-1.02, 0.48  \\
1 & -0.8 & -0.8 &  LO& LO &-0.47, 0.29  \\
1 & -0.8 & -0.8 &  TR& TR &-2.36, 0.90  \\
1 & -0.8 & -0.8 &  LO& TR &-0.61, 0.33  \\
1 & -0.8 & -0.8 &  TR& LO &-0.89, 0.47  \\
\hline
  -1 & 0.8 & -0.8 & TR+LO & TR+LO &-0.84, 0.46  \\
-1 & 0.8 & -0.8 & LO & TR+LO &-0.60, 0.39  \\
-1 & 0.8 & -0.8 & TR & TR+LO &-1.20, 0.65  \\
-1 & 0.8 & -0.8 & TR+LO & LO &-0.76 0.57  \\
-1 & 0.8 & -0.8 & TR+LO & TR &-1.10, 0.56  \\
-1 & 0.8 & -0.8 & LO & LO &-0.70, 0.64  \\
-1 & 0.8 & -0.8 & TR & TR &-2.36, 0.98  \\
-1 & 0.8 & -0.8 & LO & TR &-0.68, 0.40  \\
-1 & 0.8 & -0.8 & TR & LO &-0.90, 0.65  \\
\end{tabular}
\end{ruledtabular}
\end{table}

\begin{table}
\caption{Same as Table II but for $\sqrt{s}=1$ TeV. \label{tab3}}
\begin{ruledtabular}
\begin{tabular}{cccccc}
$\lambda_{0}$&$\lambda_{e}$ & $P_{e} $& $\lambda_{W}$&
 $\lambda_{Z}$& $a_{n}$ \\
\hline
1 & -0.8 & -0.8 &  TR+LO& TR+LO &-0.08, 0.07  \\
1 & -0.8 & -0.8 &  LO& TR+LO &-0.05, 0.04  \\
1 & -0.8 & -0.8 &  TR& TR+LO &-0.17, 0.13  \\
1 & -0.8 & -0.8 &  TR+LO& LO &-0.07, 0.06  \\
1 & -0.8 & -0.8 &  TR+LO& TR &-0.15, 0.11  \\
1 & -0.8 & -0.8 &  LO& LO &-0.03, 0.03  \\
1 & -0.8 & -0.8 &  TR& TR &-0.62, 0.33  \\
1 & -0.8 & -0.8 &  LO& TR &-0.08, 0.06  \\
1 & -0.8 & -0.8 &  TR& LO &-0.12, 0.10  \\
\hline
-1 & 0.8 & -0.8 & TR+LO & TR+LO &-0.11, 0.08  \\
-1 & 0.8 & -0.8 & LO & TR+LO &-0.06, 0.05  \\
-1 & 0.8 & -0.8 & TR & TR+LO &-0.18, 0.14  \\
-1 & 0.8 & -0.8 & TR+LO & LO &-0.10, 0.09  \\
-1 & 0.8 & -0.8 & TR+LO & TR &-0.16, 0.11  \\
-1 & 0.8 & -0.8 & LO & LO &-0.06, 0.06  \\
-1 & 0.8 & -0.8 & TR & TR &-0.63, 0.33  \\
-1 & 0.8 & -0.8 & LO & TR &-0.08, 0.06  \\
-1 & 0.8 & -0.8 & TR & LO &-0.13, 0.11  \\
\end{tabular}
\end{ruledtabular}
\end{table}

\end{document}